\documentclass[a4paper]{article}

\usepackage{icrc2013}
\usepackage[english]{babel}

\title{FlashCam: A fully digital camera for the Cherenkov Telescope Array}

\shorttitle{FlashCam: A fully digital camera for CTA}

\authors{
G.~P\"uhlhofer$^{2}$,
C.~Bauer$^{1}$, 
F.~Eisenkolb$^{2}$, 
D.~Florin$^{3}$, 
C.~F\"ohr$^{1}$, 
A.~Gadola$^{3}$, 
G.~Hermann$^{1}$, 
C.~Kalkuhl$^{2}$, 
J.~Kasperek$^{4}$, 
T.~Kihm$^{1}$, 
J.~Koziol$^{5}$, 
A.~Manalaysay$^{3}$, 
A.~Marszalek$^{5}$, 
P.J.~Rajda$^{4}$, 
W.~Romaszkan$^{4}$, 
M.~Rupinski$^{4}$, 
T.~Schanz$^{2}$, 
S.~Steiner$^{3}$, 
U.~Straumann$^{3}$, 
C.~Tenzer$^{2}$, 
A.~Vollhardt$^{3}$, 
Q.~Weitzel$^{1}$, 
K.~Winiarski$^{4}$, 
K.~Zietara$^{5}$,
for the CTA consortium$^{6}$.
}

\afiliations{
$^1$ Max-Planck-Institut f\"ur Kernphysik, P.O. Box 103980, D 69029 Heidelberg, Germany \\
$^2$ Institut f\"ur Astronomie und Astrophysik, Abteilung Hochenergieastrophysik, Kepler Center for Astro and Particle Physics, Eberhard-Karls-Universit\"at, Sand 1, D 72076 T\"ubingen, Germany \\
$^3$ Physik-Institut, Universit\"at Z\"urich, Winterthurerstrasse 190, 8057 Z\"urich, Switzerland \\
$^{4}$ AGH University of Science and Technology, Al. Mickiewicza 30, 30-059 Krakow, Poland \\
$^{5}$ Jagiellonian University, Golebia 24, 31-007 Krakow, Poland\\
$^{6}$ See http://www.cta-observatory.org/?q=node/22 for full author \& affiliation list
}

\email{Gerd.Puehlhofer@astro.uni-tuebingen.de}

\abstract{FlashCam is a Cherenkov camera development project centered around a fully digital trigger and readout scheme with smart, digital signal processing, and a ``horizontal'' architecture for the electromechanical implementation. The fully digital approach, based on commercial FADCs and FPGAs as key components, provides the option to easily implement different types of triggers as well as digitization and readout scenarios using identical hardware, by simply changing the firmware on the FPGAs. At the same time, a large dynamic range and high resolution of low-amplitude signals in a single readout channel per pixel is achieved using compression of high amplitude signals in the preamplifier and signal processing in the FPGA. The readout of the front-end modules into a camera server is Ethernet-based using standard Ethernet switches. In its current implementation, data transfer and backend processing rates of $\sim$3.8 GBytes/sec have been achieved. Together with the dead-time-free front end event buffering on the FPGAs, this permits the cameras to operate at trigger rates of up to several tens of kHz.

In the horizontal architecture of FlashCam, the photon detector plane (PDP), consisting of photon detectors, preamplifiers, high voltage-, control-, and monitoring systems, is a self-contained unit, which is interfaced through analogue signal transmission to the digital readout system. The horizontal integration of FlashCam is expected not only to be more cost efficient, it also allows PDPs with different types of photon detectors to be adapted to the FlashCam readout system. This paper describes the FlashCam concept, its verification process, and its implementation for a 12 m class CTA telescope with PMT-based PDP.}

\keywords{CTA, Imaging Atmospheric Cherenkov Telescopes, gamma-rays, electronics.}

\begin{document}
\maketitle

\section{The FlashCam concept and camera architecture}

Within the CTA consortium \cite{bib:Design2011}, the FlashCam team pursues the goal of providing a camera readout system which is based on purely digital processing of the photosensor signals. The basic concept is simple and straightforward. The charge signals from the photodetectors are digitized using commercial Flash ADCs. Driven by the telecommunication market, such devices are meanwhile available at low cost with the necessary sampling frequencies and low power consumption. The digitized pixel signals are then fed into low-cost FPGAs, for triggering, signal processing, and event transmission. 

Cherenkov telescope cameras normally run in self-triggered mode to detect air shower events. In the FlashCam concept, the trigger decision is derived digitally while the continuously sampled pixel information is temporarily stored in the FPGA. Therefore, there is no need for separate trigger electronics splitting off their signals from each preamplified pixel line. A camera-wide event trigger decision is derived using digital communication between the readout units. Such a digital trigger is very accurate and scalable, and the trigger logic is programmable even during run-time of the experiment. 

Digital preprocessing of the pixel information in the FPGA also permits to provide the full dynamic range per pixel without having to physically split the pixel signal into a low-gain and a high-gain channel. This saves a factor two in digitization units. To be able to do so, the preamplifier has two distinct response regimes, a linear (high-gain) and a logarithmic (low-gain) range. In the FPGA, the signal is simultaneously processed both for the linear range, performing pulse-shape analysis, and summing up the signal over a predefined time window to provide accurate response in the logarithmic range. Thus, a pair of ``virtual'' high- and low-gain channels per pixel are provided. Like the trigger logic, also the readout scheme is programmable through exchanging the FPGA firmware, during the livetime of the experiment, without the need of physically accessing the camera.

Data transmission from the front-end to a camera server is performed using a high-performance Ethernet network protocol. Using this Raw-Ethernet protocol, 10 Gbit links can transmit several GBytes per second. This permits to run the camera dead-time free at rates of several tens of kHz. The camera can run in self-triggered mode, transmitting all triggered events, or wait for an array trigger derived from a coincidence over several telescopes before sending the event data.

The FlashCam concept comprises a ``horizontal'' integration of its components. The photon detector plane (PDP), which consists of photon detectors, preamplifiers, high voltage-, control-, and monitoring systems, is an isolated unit. It connects to the signal digitization and triggering electronics through analog transmission cables only. This readout electronics is organized in boards which are kept in ``mini-crates''. Besides the possibility to cool the components with straightforward methods, the concept also allows adaption of various photon detectors and pixel pitches. It also reduces weight directly at the focal plane, and results in cost savings compared to other approaches.

\begin{figure}[t]
  \centering
  \includegraphics[width=0.43\textwidth]{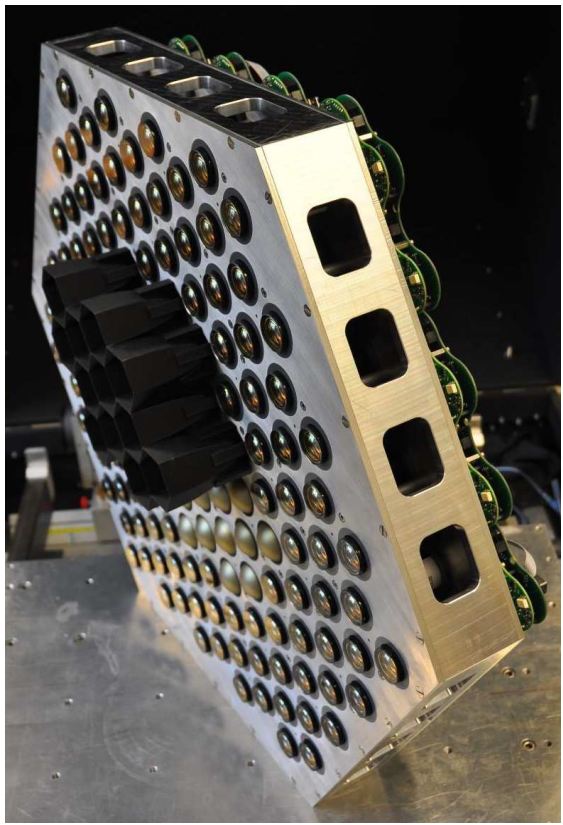}
  \caption{Photon detector plane panel of the FlashCam 144-pixel setup. For demonstrating purposes, one 12-pixel PDP module has been equipped with a fake light-guide unit. PDP modules are plugged in from behind. }
  \label{fc_pdp144_front}
\end{figure}

\begin{figure}[t]
  \centering
  \includegraphics[width=0.47\textwidth]{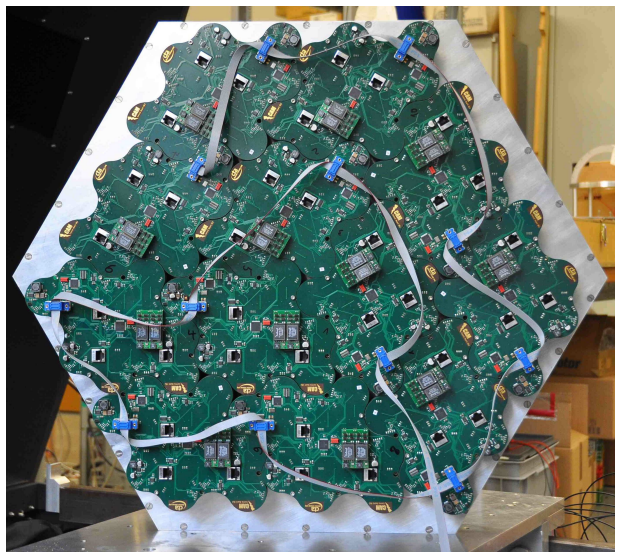}
  \caption{The 144-pixel PDP panel from behind. PDP modules are wired with the CAN-bus control cable. In addition, each 12-pixel PDP module has three CAT5/6 sockets for the analogue transmission cables.}
  \label{fc_pdp144_back}
\end{figure}

\section{Concept verification}
During the past two years, the core concepts of FlashCam have been verified using extensive simulations and demonstrator hardware. PDP modules using CTA PMTs (Hamamatsu R11920) and the FlashCam preamplifier design were used together with demonstrator FADC readout boards to verify the pixel time and amplitude resolution against simulations and CTA requirements. Key specifications are an amplitude resolution better than CTA requirements, with a dynamic range from single photoelectron resolution to over 3000 photoelectrons, and a time resolution of $<$1 nsec for signals $>$5 photoelectrons.

The demo boards and FPGA simulations were also used to verify that the envisaged low-cost Xilinx Spartan 6-FPGA provides enough processing power to cope with the required amount of data. Extensive tests were done to verify the performance of the analog signal cables (CAT 5/6) transmitting the preamplified signals from the PDP to the readout boards, including connectors and plugs. Also, extensive simulations were performed to determine the optimum baseline trigger processing algorithms and the corresponding required bandwidth. Transmission setups showed that the necessary digital trigger bandwidth can indeed be achieved. 

Data transmission setups using the raw Ethernet protocol and standard network hardware were used to verify that infrastructure for the required data rates can be provided; 3.8 GByte/sec using four\footnote{For lower readout rates, the number of Ethernet interfaces can be reduced according to the required maximum event rates.}  10 GBit interfaces have meanwhile been achieved. A 1764 pixel 12-meter telescope-class camera can be run with a readout rate of up to 36 kHz, including data processing on the camera server with array trigger and event building. Without compression, data for an event rate of up to 16 kHz can be stored to disk or sent to data management stages via an additional 10 GBit Ethernet interface.

More details of the concept and verification can also be found in \cite{bib:Gamma2012} and \cite{bib:ICRC2011}.

\begin{figure}[t]
  \centering
  \includegraphics[width=0.42\textwidth]{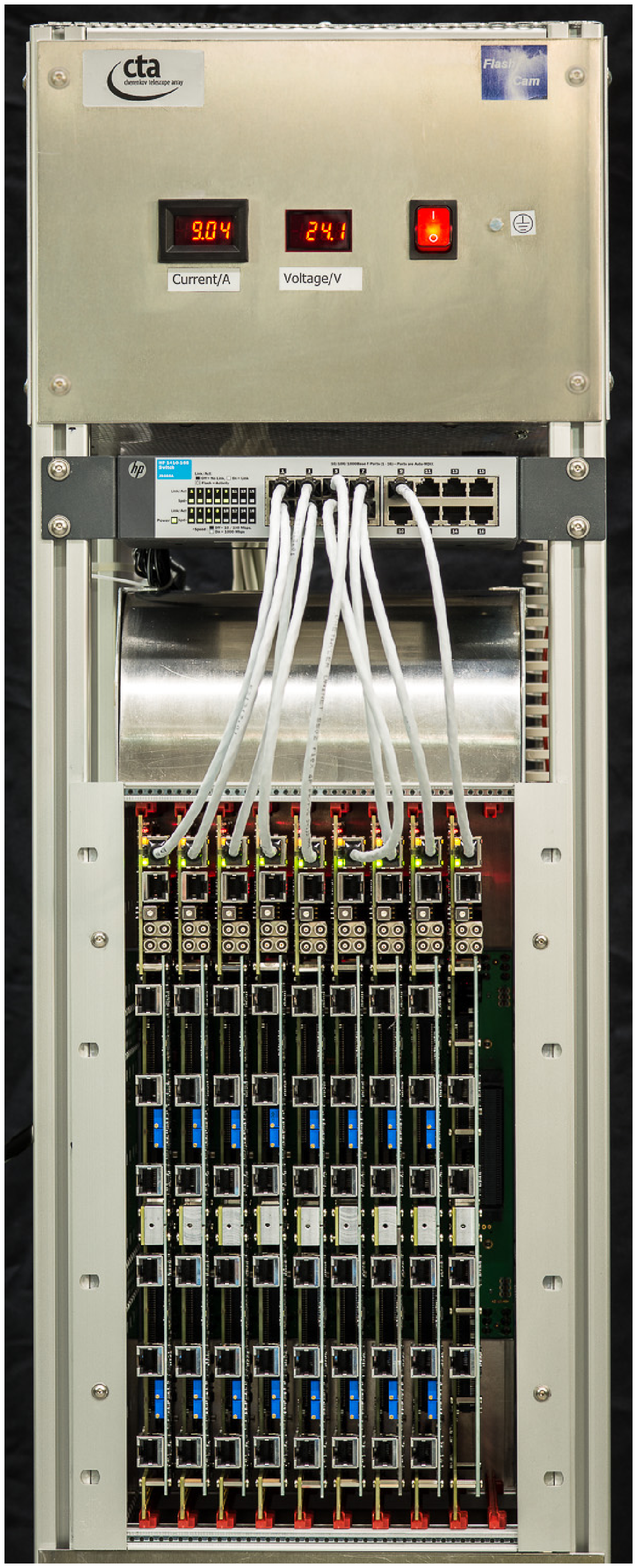}
  \caption{One fully-equipped FlashCam mini-crate test setup, holding eight FADC readout boards and one trigger distribution board (on the right). The 1 Gbit Ethernet sockets have been connected to an Ethernet switch. Other CAT 5/6 sockets take the analogue signal cables coming from the PDP modules (six per FADC board). One socket per board is for the CTI. Trigger signals are routed from the FADC boards via the backplane to the trigger distribution board.
Between mini-crate and switch, there is a deflector plate for air cooling. At the upper end of the rack, a power supply unit is mounted.  
  }
  \label{fc_minicrate}
\end{figure}

\begin{figure}[t]
  \centering
  \includegraphics[width=0.4\textwidth]{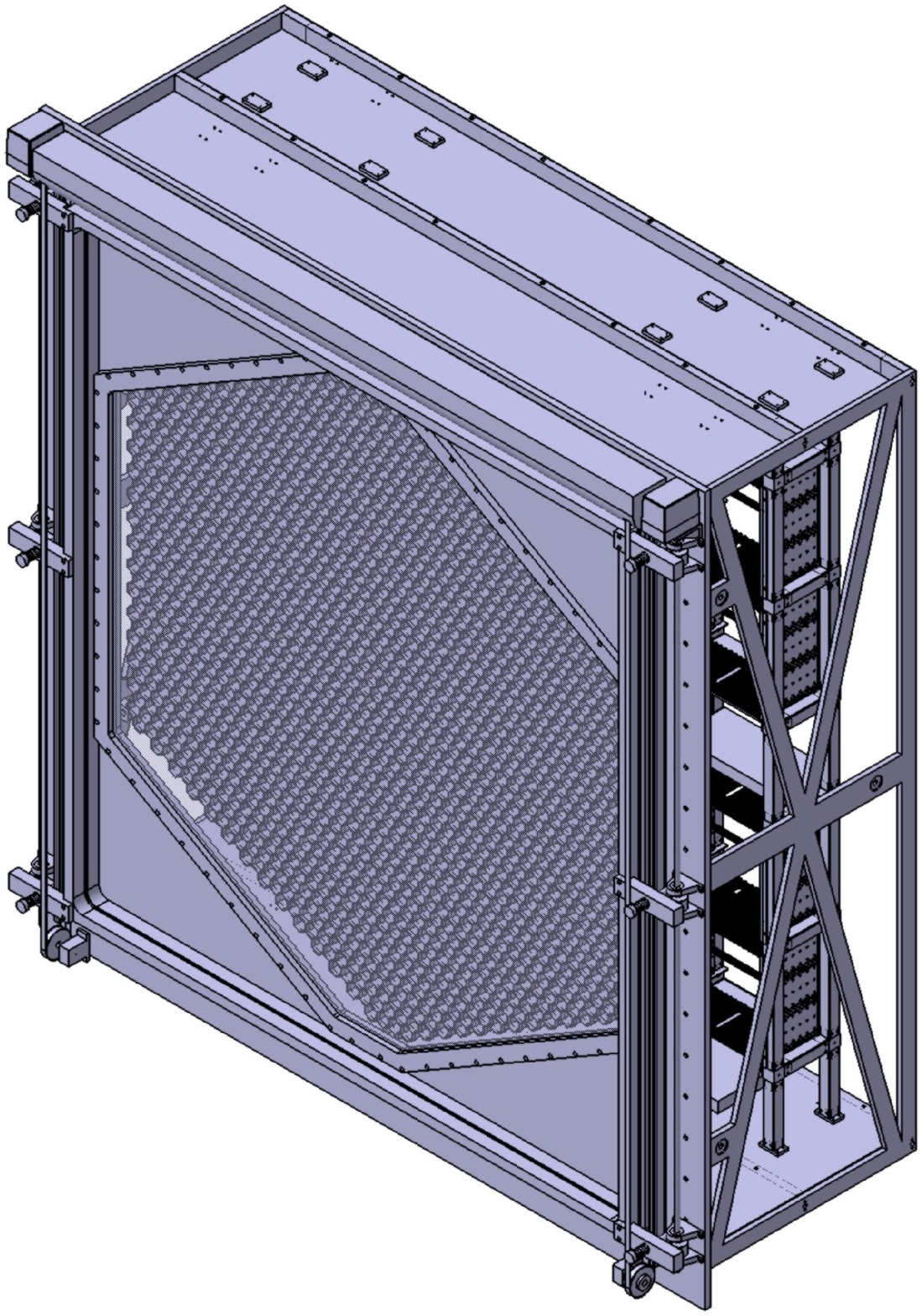}
  \includegraphics[width=0.4\textwidth]{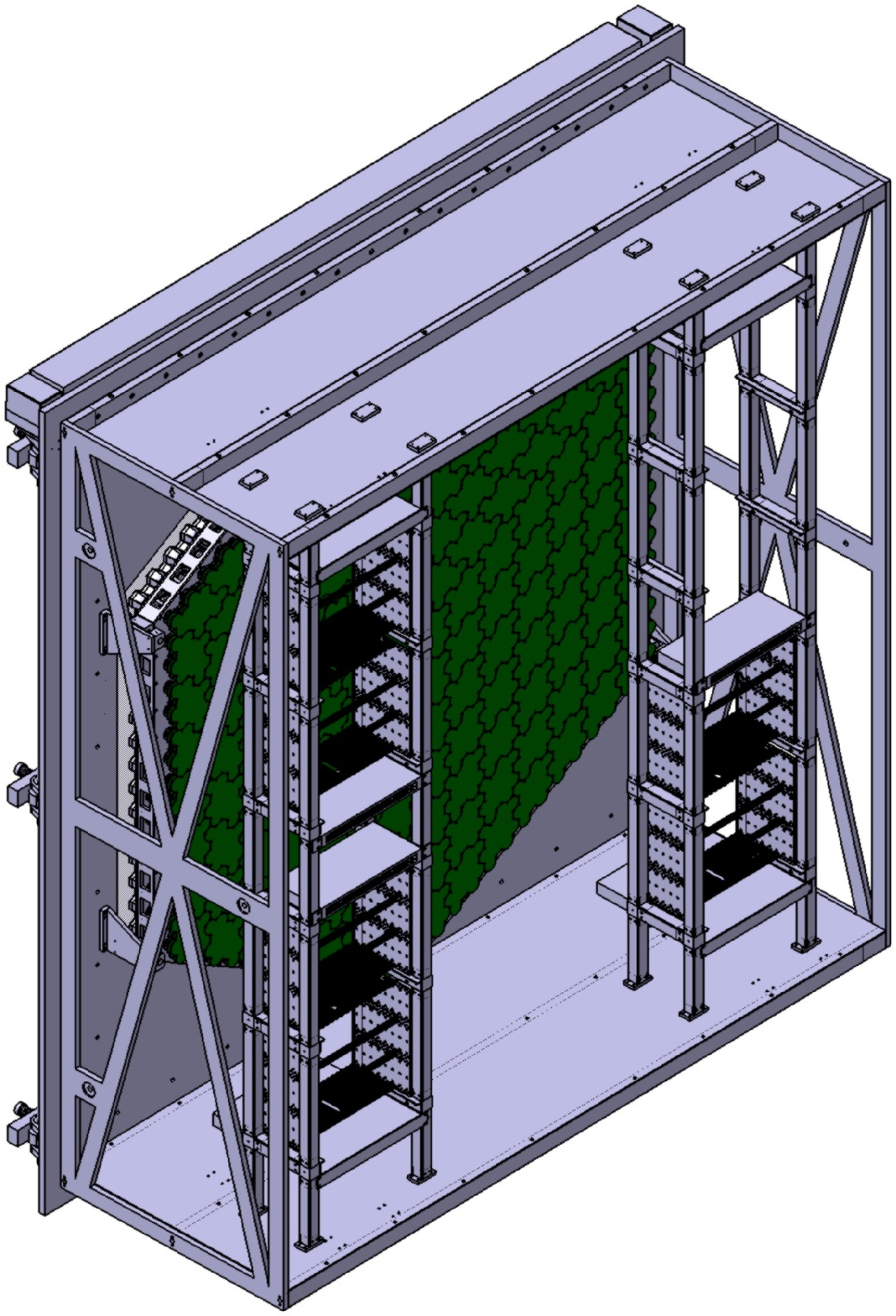}
  \caption{Drawings of a 1764-pixel FlashCam camera using PMTs. The front lid of the camera can be protected by a roller shutter. The racks in the camera back take the readout-electronics crates. The PDP modules are plugged into the front panel from behind, i.e. from inside the camera body, without the need for removing the front-lid and light guides.}
  \label{fc_camera_front}
\end{figure}

\section{The photon detector and electronics building blocks}
A major challenge for any type of possible CTA camera is the engineering-wise clean and solid implementation of the camera electronics. This is why the FlashCam team has decided to build and test the major building blocks (PDP modules and mini-crates) as soon as possible, to demonstrate the concept on relevant camera scales and to prepare for mass production, but also to understand in which areas improvement on the engineering level might be needed, and to still have enough time to implement such improvements within the CTA schedule for prototyping.

For the PDP, modules of 12 pixels were developed; the number provides the best compromise between high integration and flexibility for the focal plane geometry. Each module contains PMTs, preamplifiers, HV supply, and a microcontroller for slow control and monitoring. CAN-bus control and 24 V power supply are provided through a 9-pin D-sub connector. Low and high voltage are generated on the module. Pixels are individually controllable (HV off, HV -700 V .. -1500 V).

The readout electronics boards are kept in crates and feature a higher -- compared to the PDP boards -- integration of 24 channels (pixels) per board. In fact, all readout boards are based on identical motherboards equipped with a Spartan-6 FPGA. The functionality of the boards is defined by plugging in mezzanine (``daughter'') boards. To create digitization boards, motherboards are equipped with FADC mezzanine cards. In addition to the currently used card with FADC chips from Texas Instruments, an alternative board with chips from Analog Devices is also being developed, 
to retain some flexibility in the choice of the supplier for mass production. Both chips sample at 250 MHz frequency, which was shown during the concept veri\-fi\-cation to deliver the required trigger and pixel resolution performance, when incorporated in the FlashCam concept. Trigger interface and clock/master daughter boards have also been developed. 

The expected data rate of the system allows the full pixel event information to be transmitted over standard gigabit Ethernet infrastructure without any data reduction. In the current geometry, up to 96 FADC boards can be accommodated for, corresponding to 2304 pixels, sufficient for all telescope types envisaged within CTA. All FADC boards have a 1 Gbit Ethernet connection each, which will be routed through one or two 10 Gbit Ethernet lines for data transmission to the camera server.

\section{Mini-camera setup}

After verification of the individual components, the FlashCam team is preparing a 144-pixel setup to verify the functionality of the system and interplay between the components on a full camera-relevant size scale. Fig. \ref{fc_pdp144_front} and \ref{fc_pdp144_back} show the PDP setup with 12 PDP boards, which is fully tested and functional. The sandwich mechanics which represents a scaled-down version of the currently foreseen PDP mechanics was verified to conform to the stability expectations as derived from FEM calculations.

Fig. \ref{fc_minicrate} shows a fully equipped mini-crate, featuring eight FADC boads and one trigger distribution board. The anticipated limit of 1.3 W per pixel has been achieved. Individual boards have been verified for full functionality, trigger communication verification tests between the boards are ongoing. After those, the PDP 144-pixel panel and an electronics setup will be evaluated in conjunction, the latter featuring all connectivity which is relevant to demonstrate the functionality of a full-scale camera.

\section{Towards a full-scale prototype}

The FlashCam team is currently preparing the equipment for a full-scale camera prototype. Drawings of the prototype body are shown in Fig. \ref{fc_camera_front}. The readout electronics for PMT-based cameras will be mounted in the camera bodies. The camera will have a weight of significantly less than 2 tons, well below the requirement limit of 2.5 tons to which the telescope structure is designed for.

The main challenge comes from cooling. The camera body will be nearly air-tight, with slight over-pressure inside, to prevent dust from entering. The total power dissipation will be of the order of 4.5 kW for a 1764-pixel camera. The PDP creates about 400 W of heat, and likely does not need special cooling. The largest fraction of the power ($\sim$ 2.6 kW) will be produced by the readout electronics in the crates. The current concept foresees cooling with fans, using cooled airflow. Options how to provide such airflow using water-air heat exchangers are part of the ongoing evaluations towards a full-scale FlashCam camera prototype.

The separation of PDP and readout electronics permits to install and exchange PDP modules (including PMTs) from inside the camera, i.e. without having to remove the transparent camera front-lid and the light guides which concentrate the incoming light onto the PMT photocathodes. 
The camera body space also permits to install and exchange individual electronics boards at the mini-crates by a human entering the camera. When not operational, the front-lid will be protected by a roller shutter.

\section{Conclusion}

The FlashCam team has demonstrated that a fully-digital camera readout concept is suitable for classical PMT-based Cherenkov telescope cameras. Full verification of the concept against simulations and CTA requirements has been performed. All necessary components for a PMT-based camera have been developed. An intermediate, 144-pixel FlashCam mini-camera is currently being set up to demonstrate functionality of the system on camera-relevant size scales. A full-scale prototype camera for the 12-meter medium-sized CTA telescope is currently being prepared. At the same time, the readout concept is also well-matched for other photosensors such as GAPDs, as being pursued by the CTA one-mirror small size telescope project \cite{bib:SSTICRC2013}.

%\section*{Acknowledgements}

\vspace*{0.5cm}
\footnotesize{{\bf Acknowledgment:}{We gratefully acknowledge support from the agencies and organisations listed in this page: 
http://www.cta-observatory.org/?q=node/22.}}

\end{document}